\documentclass[12pt]{article}

\begin{document}

\begin{center}

{\Large {UV-finite "old" conformal bootstrap on AdS:
\\
 scalar case}}\footnote{{\it JHEP} {\bf 01} (2020) 137}

\vspace{1,5cm}

{Boris L. Altshuler}\footnote{E-mail addresses: baltshuler@yandex.ru $\,\,\,  \& \,\,\,$  altshul@lpi.ru}

\vspace{1cm}

{\it Theoretical Physics Department, P.N. Lebedev Physical
Institute, \\  53 Leninsky Prospect, Moscow, 119991, Russia}

\vspace{1,5cm}

\end{center}

{\bf Abstract:} The double-trace from UV to IR flow subtraction of infinities used earlier for the UV-convergent calculations of the Witten tadpole diagrams being applied to the bubble self-energy diagrams gives for them the amazingly simple expressions in case of the four-dimensional boundary space. For every $N = 1... 4$ in the $O(N)$ symmetric scalar fields model with the conformal Hubbard-Stratonovich field there are three roots of the "old" conformal bootstrap spectral equations that obey unitarity bound demand.

\vspace{0,5cm}


\vspace{0,5cm}


\newpage

\tableofcontents

\newpage

\section{Introduction}

\quad Masses of many particles of Standard Model are below the SM scale and their prediction remains one of the main challenges of theoretical physics. In frames of the AdS/CFT approach and Randall-Sundrum model \cite{Randall} spectra of physical particles are obtained as eigenvalues of equations for bulk fields, and it is possible in principle to get the looked for masses of intermediate scale with the choice of bulk masses of the fields.

In papers \cite{Alt0} it was shown that "old" conformal bootstrap (proposed about 50 years ago in pioneer papers \cite{old1}, \cite{old2} and developed in \cite{Parisi} - \cite{Dobrev2}, see e.g. \cite{Grensing} and references therein) considered in the AdS/CFT context permits to calculate conformal dimensions, that is bulk masses of the fields. The simplest "old" conformal bootstrap equation for Green function $G(X, Y)$ traditionally written in planar approximation looks as: 

\begin{equation}
\label{1}
G (X_{1}, X_{2}) = g^{2} \, \int\int\, G(X_{1}, X)\,G(X, Y) \, G(X, Y) \, G(Y, X_{2})\, dX\,dY
\end{equation}
(triple interaction is supposed, $g$ is the coupling constant). Equating of exact Green functions to the one-loop quantum contribution built of the same exact Green functions is the main postulate of the "old" conformal bootstrap. Schwinger-Dyson Eq-s of type (\ref{1}) with "tadpole" self-energy and contribution from massless "bare" Lagrangian in the RHS are used in theories with dynamical generation of mass (gap) such as superconductivity and certain models of spontaneous symmetry breaking. In some models the choice of only planar diagrams in the RHS of (\ref{1}) may be justified as "most divergent" ones or in frames of the $1/N$ expansion. However, Kenneth Wilson in his 1982 Nobel Lecture \cite{Wilson} criticized this approximation as ungrounded. Nevertheless people use it, and we shall do the same in the present paper, having in mind that interesting results are the best justification of any postulate. 

To see the meaning of Eq. (\ref{1}) in the AdS/CFT context let us assume that $X_{1,2}$, $X$, $Y$ in (1) are the bulk coordinates in $AdS_{d + 1}$ and direct $X_{1, 2}$ to horizon (AdS boundary at $z_{0} \to 0$ in Poincare coordinates). Then LHS of (\ref{1}) becomes conformal correlator of the boundary conformal theory, whereas $G(X_{1}, X)$, $G(Y, X_{2})$ in the RHS become the corresponding bulk-to-boundary propagators. Thus RHS of (\ref{1}) becomes the quantum one-loop self-energy contribution to the boundary-to-boundary correlator; this Witten diagram is called "bubble" \cite{Giombi1}. 

The spectra of conformal dimensions obtained in \cite{Alt0} were calculated under the simplifying assumption that Green functions that form a bubble in the RHS of (\ref{1}) may be replaced by the corresponding harmonic (Wightman) functions. Here we abandon this assumption.

It is well known that one-loop diagrams in the RHS of (\ref{1}) are plagued by UV divergencies. In particular it is seen in the divergence of the double-integral spectral representation of the bubble \cite{Giombi1}.
To overcome this difficulty we propose to apply to Witten bubble diagrams the double-trace from UV to IR flow approach used in \cite{Mitra} - \cite{Diaz2} for UV finite calculations of tadpoles and quantum vacuum energies of scalar and spinor bulk fields in spaces of arbitrary dimensions. This "flow" is just a difference of two similar Witten diagrams built of the UV or IR Green functions, and it proves to be finite and well defined. 

We propose to apply this approach to Witten bubble diagrams. Most generally this approach means that instead of standard quantum generation functional (symbolically)

\begin{equation}
\label{2}
Z [j;G] = ({\rm{Det}}G)^{-1/2}\, e^{L_{int}\left(\frac{\delta}{\delta j}\right)} \, e^{\left(\frac{1}{2}jGj\right)}
\end{equation}
(here $G$, $L_{int}$, $j$ are the free field Green function, the interaction Lagrangian and field's source) the ratio

\begin{equation}
\label{3}
{\widetilde Z} [j;G^{UV}, G^{IR}] = \frac{Z [j;G^{UV}]}{Z [j;G^{IR}]} = \frac{({\rm{Det}}G^{UV})^{-1/2}\, e^{L_{int}\left(i\frac{\delta}{\delta j}\right)} \, e^{\left(\frac{1}{2}jG^{UV}j\right)}}{({\rm{Det}}G^{IR})^{-1/2}\, e^{L_{int}\left(i\frac{\delta}{\delta j}\right)} \, e^{\left(\frac{1}{2}jG^{IR}j\right)}}
\end{equation}
of two quantum functionals determined by Green functions ($G^{UV}$ and $G^{IR}$) possessing two different asymptotics at the AdS boundary must be considered as quantum generation functional for Witten diagrams. This means in particular that self-energy in the RHS of (\ref{1}) is defined as a difference of conventional Witten diagrams built from the products of two $G^{UV}$ and two $G^{IR}$ Green functions correspondingly. Then the diverging double spectral integrals are canceled in $(G^{UV})^{2} - (G^{IR})^{2}$ and the remaining terms are UV-finite, see it in Sec. 3.

In Sec.2 familiar expressions used in the bulk of the paper are summed up. In Sec. 3 the expression for the UV-finite one-loop self-energy correlator is obtained that proves to be surprisingly simple in $AdS_{5}$ that is for $d = 4$. This is one of the main results of the paper. In Sec. 4 "old" conformal bootstrap equations in the AdS/CFT context are written down for the $O(N)$ symmetric model of $N$ scalar fields $\psi_{i}$ of one and the same conformal dimension $\Delta_{\psi}$ interacting with the Hubbard-Stratonovich conformal invariant auxiliary scalar field. "Old" conformal bootstrap gives non-trivial spectral equation for $\Delta_{\psi}$. This equation as well as calculation of the roots of this equation obeying unitarity bound demand is another result of the paper. In Conclusion the possible directions of future work are outlined.

\section{Preliminaries}

\quad We work in $AdS_{d+1}$ in Poincare Euclidean coordinates $Z = \{z_{0}, {\vec z}\,\}$, where AdS curvature radius $R_{AdS}$ is put equal to one:

\begin{equation}
\label{4}
ds^{2} = \frac{dz_{0}^{2} + d {\vec z}\,^{2}}{z_{0}^{2}},
\end{equation}
and consider bulk scalar fields. Bulk field $\phi (X)$ of mass $m$ is dual to boundary conformal operator $O_{\Delta^{IR}} (\vec x)$ or to its "shadow" operator $O_{\Delta^{UV}}(\vec x)$ with scaling dimensions

\begin{equation}
\label{5}
\Delta_{\phi}^{IR} = \frac{d}{2} + \sqrt{\frac{d^{2}}{4} + m^{2}} > \frac{d}{2}, \, \, \, \,  \Delta_{\phi}^{UV} = d - \Delta_{\phi}^{IR} < \frac{d}{2}.
\end{equation}

We take normalization of the scalar field bulk-to-boundary propagator $G^{\partial B}_{\Delta} (Z; \vec x)$ and of the corresponding conformal correlator like in \cite{Giombi1}:

\begin{eqnarray}
\label{6}
G^{\partial B}_{\Delta} (Z; \vec x) = \lim_{\stackrel {x_{0} \to 0}{}} \left[\frac{G_{\Delta}^{BB} (Z, X)}{(x_{0})^{\Delta}}\right] = C_{\Delta}\, \left [\frac{z_{0}}{z_{0}^{2} + (\vec z - \vec x)^{2}}\right]^{\Delta}, \nonumber
\\
\\
 C_{\Delta} = \frac{\Gamma (\Delta)}{2\pi^{d/2}\Gamma \left(1 + \Delta - \frac{d}{2}\right)}, \qquad  \qquad \qquad  \qquad \nonumber
\end{eqnarray}
and:

\begin{equation}
\label{7}
<O_{\Delta}({\vec x}) O_{\Delta} ({\vec y})> = \lim_{\stackrel{x_{0} \to 0}{y_{0} \to 0}} \left[\frac{G_{\Delta}^{BB} (X, Y)}{(x_{0}\,y_{0})^{\Delta}}\right]= \frac{C_{\Delta}}{P_{xy}^{\Delta}}, \, \, \,  P_{xy} = |{\vec x} - {\vec y}|^{2}
\end{equation}

Bulk-to-bulk IR ($\Delta = \Delta^{IR}$) scalar field Green function $G_{\Delta}^{IR} (X, Y)$ that is zero at infinity $x_{0}, y_{0} \to \infty$ possesses Kallen-Lehmann type spectral representation \cite{Penedones} - \cite{Bekaert}, \cite{Giombi1}:

\begin{equation}
\label{8}
G^{IR}_{\Delta} (X, Y) = \int_{-\infty}^{+\infty} \frac{\Omega_{\nu,0}(X,Y)\,d\nu}{[\nu^{2} + (\Delta - \frac{d}{2})^{2}]}, \qquad  \qquad  \qquad \qquad
\end{equation}
where nominator of the integrand is scalar field Harmonic function that admits split representation and that is proportional to the difference (marked here with tilde) of IR and UV bulk Green functions:

\begin{equation}
\label{9}
\Omega_{\nu,0}(X,Y) = \frac{\nu^{2}}{\pi} \, \int \, G^{\partial B}_{\frac{d}{2} + i\nu}(X, {\vec x}_{a})\, G^{\partial B}_{\frac{d}{2} - i\nu}(Y, {\vec x}_{a})\, d^{d}{\vec x}_{a}  = \frac{i\nu}{2\pi}\,{\widetilde G}_{\frac{d}{2} + i\nu}, \qquad
\end{equation}

\begin{equation}
\label{10}
{\widetilde G}_{\Delta}(X, Y) = G_{\Delta}^{IR} - G_{d-\Delta}^{UV} = (d - 2\Delta)\, \int \, G^{\partial B}_{\Delta}(X, {\vec x}_{a})\, G^{\partial B}_{d-\Delta}(Y, {\vec x}_{a})\, d^{d}{\vec x}_{a}.
\end{equation}

Spectral representation for $G^{UV}_{d - \Delta}$ was analyzed in detail in \cite{Giombi1}, but after all it comes to the identity:

\begin{equation}
\label{11}
G^{UV}_{d - \Delta}(X, Y) = G^{IR}_{\Delta}(X, Y) - {\widetilde G}_{\Delta}(X, Y),
\end{equation}
where $G^{IR}_{\Delta}(X, Y)$ and ${\widetilde G}_{\Delta}(X, Y)$ are given in (\ref{8}) and (\ref{10}).

We shall also need expression for AdS/CFT tree 3-point vertex \cite{Freedman}, \cite{Penedones}, \cite{Paulos}, \cite{Giombi1}:

\begin{eqnarray}
\label{12}
\Gamma_{\Delta_{1}, \Delta_{2}, \Delta_{3}} ({\vec x}_{1}, {\vec x}_{2}, {\vec x}_{3}) = 
\int \,G^{\partial B}_{\Delta_{1}} (X; {\vec x}_{1})\, G^{\partial B}_{\Delta_{2}} (X; {\vec x}_{2}) \, G^{\partial B}_{\Delta_{3}} (X; {\vec x}_{3}) \, dX =  \nonumber
\\  
\\
= \, \frac{B(\Delta_{1}, \Delta_{2}, \Delta_{3})}{P_{12}^{\delta_{12}}\,P_{13}^{\delta_{13}}\, P_{23}^{\delta_{23}}}, \qquad  \, \qquad \, \qquad  \, \qquad \, \qquad  \nonumber
\end{eqnarray}
where

\begin{equation}
\label{13}
\delta_{12} = \frac{\Delta_{1} + \Delta_{2} - \Delta_{3}}{2}; \,\,\, \delta_{13} = \frac{\Delta_{1} + \Delta_{3} - \Delta_{2}}{2}; \,\,\, \delta_{23} = \frac{\Delta_{2} + \Delta_{3} - \Delta_{1}}{2},
\end{equation}

\begin{equation}
\label{14}
B(\Delta_{1}, \Delta_{2}, \Delta_{3}) = \frac{\pi^{d/2}}{2}\, \left( \prod\limits_{i=1}^{3}\frac{C_{\Delta_{i}}}{\Gamma (\Delta_{i})}\right) \cdot \Gamma \left(\frac{\Sigma \Delta_{i} - d}{2}\right)\cdot \Gamma(\delta_{12})\, \Gamma(\delta_{13}) \, \Gamma (\delta_{23}).
\end{equation}

Also some well known \cite{Symanzik2}, \cite{Parisi}, \cite{Fradkin}, \cite{Giombi1} conformal integrals will be used below:

\begin{equation}
\label{15}
\int \frac{d^{d}{\vec y}}{P_{1y}^{\Delta_{1}} \,P_{2y}^{\Delta_{2}}\, P_{3y}^{\Delta_{3}}} \stackrel {\Sigma \Delta_{i} = d} {=} \frac{A(\Delta_{1}, \Delta_{2}, \Delta_{3})}{P_{12}^{\frac{d}{2} - \Delta_{3}}\,P_{13}^{\frac{d}{2} - \Delta_{2}}\, P_{23}^{\frac{d}{2} - \Delta_{1}}},
\end{equation}
and

\begin{equation}
\label{16}
\int \frac{d^{d}{\vec y}}{P_{1y}^{\Delta_{1}} \,P_{2y}^{\Delta_{2}}} = \frac{A(\Delta_{1}, \Delta_{2}, d - \Delta_{1} - \Delta_{2})} {P_{12}^{\Delta_{1} + \Delta_{2} - \frac{d}{2}}},
\end{equation}
where

\begin{equation}
\label{17}
A(\Delta_{1}, \Delta_{2}, \Delta_{3}) = \frac{\pi^{d/2}\, \Gamma (\frac{d}{2} - \Delta_{1}) \, \Gamma (\frac{d}{2} - \Delta_{2}) \, \Gamma (\frac{d}{2} - \Delta_{3})}{\Gamma (\Delta_{1}) \, \Gamma(\Delta_{2}) \, \Gamma (\Delta_{3})}.
\end{equation}

And we shall need divergent integral (\ref{16}) when $\Delta_{1} = \Delta_{2} = d/2$:

\begin{equation}
\label{18}
\int \frac{d^{d}{\vec y}}{P_{1y}^{\frac{d}{2}} \,P_{2y}^{\frac{d}{2}}} = \frac{A(\frac{d}{2}, \frac{d}{2}, 0)}{P_{12}^{\frac{d}{2}}} = \frac{\pi^{d/2} \, \Gamma (0)}{\Gamma(\frac{d}{2})} \cdot \frac{1}{P_{12}^{\frac{d}{2}}},
\end{equation}
the possible different regularizations of (\ref{18}) are discussed in \cite{Giombi1}.

\section{UV-finite one-loop self-energy correlator}

\qquad We consider here three bulk scalar fields $\phi_{i}(Z)$ ($i = 1, 2, 3$) with conformal dimensions $\Delta_{\phi_{i}}$ (\ref{5}) and triple bulk interaction:

\begin{equation}
\label{19}
L_{int} = g\,\phi_{1}(Z) \,\phi_{2}(Z) \,\phi_{3}(Z).
\end{equation}

The RHS of bootstrap equation (\ref{1}) when $X_{1}$, $X_{2}$ are put at the horizon is the 2-point one-loop self-energy boundary-boundary correlator (bubble) ${\cal M}^{{\rm 2pt \, bub}}_{\Delta_{\phi_{1}}|\Delta_{\phi_{2}}\Delta_{\phi_{3}}}({\vec x}_{1}, {\vec x}_{2})$ built of two bulk-to-boundary propagators (\ref{6}) of the "external" field $\phi_{1}$ and of the product of two intermediate bulk Green functions of fields $\phi_{2}$ and $\phi_{3}$, this is reflected by index $\Delta_{\phi_{1}}|\Delta_{\phi_{2}}\Delta_{\phi_{3}}$.

From now on the IR option $(\Delta_{\phi_{i}} > d/2)$ will be considered when spectral representation (\ref{8}) for the intermediate Green functions is valid. The product of two bulk Green functions gives UV divergence of the bubble. Thus we $postulate$ that bubble is built with use of the "double trace from UV to IR deformation" quantum generation functional ${\widetilde Z} = Z^{UV}/Z^{IR}$ (\ref{3}) (see discussion in the Introduction), and also mark it with tilde:

\begin{eqnarray}
\label{20}
{\cal {\widetilde M}}^{{\rm 2pt \, bub}}_{\Delta_{\phi_{1}}|\Delta_{\phi_{2}}\Delta_{\phi_{3}}}({\vec x}_{1}, {\vec x}_{2}) = {\cal M}^{{\rm 2pt \, bub} \, UV}_{\Delta_{\phi_{1}}|\Delta_{\phi_{2}}\Delta_{\phi_{3}}}({\vec x}_{1}, {\vec x}_{2}) - {\cal M}^{{\rm 2pt \, bub} \, IR}_{\Delta_{\phi_{1}}|\Delta_{\phi_{2}}\Delta_{\phi_{3}}}({\vec x}_{1}, {\vec x}_{2}) = \nonumber
\\
\\
= g^{2} \, \int\int G^{\partial B}_{\Delta_{\phi_{1}}}(X ; {\vec x}_{1})\, {\widetilde \Pi}_{\Delta_{\phi_{2}}, \Delta_{\phi_{3}}}(X, Y) \, G^{\partial B}_{\Delta_{\phi_{1}}}(Y ; {\vec x}_{2})\,dXdY, \qquad  \qquad  \nonumber 
\end{eqnarray}
where

\begin{eqnarray}
\label{21}
{\widetilde \Pi}_{\Delta_{\phi_{2}}, \Delta_{\phi_{3}}}(X, Y) = \qquad  \qquad  \qquad  \qquad  \qquad \nonumber
\\
\\
= G^{UV}_{\Delta_{\phi_{2}}}(X, Y) \, G^{UV}_{\Delta_{\phi_{3}}} (X, Y) - G^{IR}_{\Delta_{\phi_{2}}}(X, Y) \, G^{IR}_{\Delta_{\phi_{3}}} (X, Y)  = \qquad \nonumber
\\ \nonumber
\\
= {\widetilde G}_{\Delta_{\phi_{2}}} \, {\widetilde G}_{\Delta_{\phi_{3}}} - G^{IR}_{\Delta_{\phi_{2}}} \, {\widetilde G}_{\Delta_{\phi_{3}}} - {\widetilde G}_{\Delta_{\phi_{2}}} \, G^{IR}_{\Delta_{\phi_{3}}},    \qquad  \qquad  \qquad \nonumber
\end{eqnarray}
last equality in (\ref{21}) is actually an identity following from the definition of ${\widetilde G_{\Delta}}$ (\ref{10}).

UV-divergent terms of $[G^{IR}]^{2}$ and $[G^{UV}]^{2}$ (that are given by the double integral spectral representations \cite{Giombi1}) reduce in (\ref{20}) whereas, as it will be shown below, terms of (\ref{20}) corresponding to the last two terms in the final line of (\ref{21}) are given by the convergent spectral integrals (\ref{8}). Since nominators of these integrals are proportional to $\widetilde G$ (see (\ref{9}), (\ref{10})) all three terms of (\ref{20})-(\ref{21}) are expressed through the one and the same correlator (we call it "harmonic bubble" ${\cal H}$), where both intermediate Green functions are replaced by their harmonic counterparts ${\widetilde G}$ (\ref{10}):

\begin{eqnarray}
\label{22}
{\cal H}^{\rm {2pt \, bub}}_{\Delta_{\phi_{1}}|\Delta_{\phi_{2}}\Delta_{\phi_{3}}}\,({\vec x}_{1}, {\vec x}_{2}) =  \qquad  \qquad \qquad  \qquad \nonumber
\\
\\
= g^{2} \int\int G^{\partial B}_{\Delta_{\phi_{1}}}(X ; {\vec x}_{1})\,{\widetilde G}_{\Delta_{\phi_{2}}}(X, Y) \, {\widetilde G}_{\Delta_{\phi_{3}}} (X, Y)\,G^{\partial B}_{\Delta_{\phi_{1}}}(Y ; {\vec x}_{2})\,dXdY.  \nonumber
\end{eqnarray}

Thus, with account of (\ref{21}), (\ref{8}), (\ref{9}), (\ref{22}) the double-trace deformation of bubble ${\widetilde M}$ (\ref{20}) takes a form:

\begin{eqnarray}
\label{23}
{\cal {\widetilde M}}^{{\rm 2pt \, bub}}_{\Delta_{\phi_{1}}|\Delta_{\phi_{2}}\Delta_{\phi_{3}}}({\vec x}_{1}, {\vec x}_{2}) = - \int_{-\infty}^{+\infty} \frac{i\nu \, d\nu}{2 \pi} \, \frac{{\cal H}^{\rm {2pt \, bub}}_{\Delta_{\phi_{1}}|\frac{d}{2} + i\nu, \Delta_{\phi_{3}}}\,({\vec x}_{1}, {\vec x}_{2})}{[\nu^{2} + (\Delta_{\phi_{2}} - \frac{d}{2})^{2}]} - \qquad \nonumber
\\
\\
- \int_{-\infty}^{+\infty} \frac{i\nu \, d\nu}{2 \pi} \, \frac{{\cal H}^{\rm {2pt \, bub}}_{\Delta_{\phi_{1}}| \Delta_{\phi_{2}}, \frac{d}{2} + i\nu}\,({\vec x}_{1}, {\vec x}_{2})}{[\nu^{2} + (\Delta_{\phi_{3}} - \frac{d}{2})^{2}]} \, + \, {\cal H}^{\rm {2pt \, bub}}_{\Delta_{\phi_{1}}| \Delta_{\phi_{2}}, \Delta_{\phi_{3}}}\,({\vec x}_{1}, {\vec x}_{2}). \qquad \nonumber
\end{eqnarray}

Harmonic bubble ${\widetilde H}$ (\ref{22}) was calculated in \cite{Giombi1}. The evident steps are as follows: (1) to use in (\ref{22}) split representations of ${\widetilde G}_{\Delta}$ (\ref{10}); (2) to perform two bulk integrals (\ref{12}) that gives in (\ref{22}) convolution of two vertices (\ref{12}) over two boundary points ${\vec x}_{a}$, ${\vec x}_{b}$; (3) to perform familiar conformal integral (\ref{15}) over ${\vec x}_{b}$. Then (\ref{22}) comes to:

\begin{eqnarray}
\label{24}
{\cal {\widetilde H}}^{{\rm 2pt \, bub}}_{\Delta_{\phi_{1}}|\Delta_{\phi_{2}}\Delta_{\phi_{3}}}\,({\vec x}_{1}, {\vec x}_{2}) = g^{2} \, (d - 2\Delta_{\phi_{2}}) (d - 2\Delta_{\phi_{3}})\,\frac{1}{P_{12}^{\Delta_{\phi_{1}} - \frac{d}{2}}}\, \int \frac{d{\vec x}_{a}}{P_{1a}^{\frac{d}{2}}P_{2a}^{\frac{d}{2}}}  \, \cdot \nonumber
\\  \nonumber
\\  \nonumber
\cdot \, B(\Delta_{\phi_{1}}, \Delta_{\phi_{2}}, \Delta_{\phi_{3}})\, B(\Delta_{\phi_{1}}, d - \Delta_{\phi_{2}}, d - \Delta_{\phi_{3}}) \, A(\delta_{12}, \delta_{13}, d - \Delta_{\phi_{1}}) = \quad \nonumber
\\
\\
= \frac{C_{\Delta_{\phi_{1}}}}{P_{12}^{\Delta_{\phi_{1}}}} \, \cdot \, \frac{g_{R}^{2}}{F(\Delta_{\phi_{1}})} \, \cdot \,  {\bf {\cal R}}(\Delta_{\phi_{1}}, \Delta_{\phi_{2}}, \Delta_{\phi_{3}}),  \qquad  \qquad \qquad \nonumber
\end{eqnarray}
$P_{12} = |{\vec x}_{1} - {\vec x}_{2}|^{2}$, etc.; standard divergent conformal integral (\ref{18}) is absorbed here, together with some coefficients, in the "bare" coupling constant $g^{2}$ defining the renormalized coupling constant as:

\begin{equation}
\label{25}
g_{R}^{2} = g^{2}\,\frac{P_{12}^{\frac{d}{2}}}{32\pi^{d}}\,\int \frac{d{\vec x}_{a}}{P_{1a}^{\frac{d}{2}}P_{2a}^{\frac{d}{2}}}.
\end{equation}
Coefficient ${\bf {\cal R}}(\Delta_{\phi_{1}}, \Delta_{\phi_{2}}, \Delta_{\phi_{3}})$ in the last line of (\ref{24}) is equal to:

\begin{eqnarray}
\label{26}
{\bf {\cal R}}(\Delta_{\phi_{1}}, \Delta_{\phi_{2}}, \Delta_{\phi_{3}}) = \Gamma\left(\frac{{\Sigma_{i}\Delta_{\phi_{i}}} - d}{2}\right) \, \Gamma\left(\frac{2d - {\Sigma_{i}\Delta_{\phi_{i}}}}{2}\right) \,  \cdot \nonumber
\\
\\
\cdot \, \frac{\Gamma(\delta_{12}) \, \Gamma(\delta_{13}) \, \Gamma(\delta_{23}) \,\Gamma\left(\frac{d}{2} - \delta_{12}\right) \, \Gamma\left(\frac{d}{2} - \delta_{13}\right) \, \Gamma\left(\frac{d}{2} - \delta_{23}\right)}{\Pi_{i=1}^{3}\left[\Gamma\left(\frac{d}{2} - \Delta_{\phi_{i}}\right) \, \Gamma\left(1 + \Delta_{\phi_{i}} - \frac{d}{2}\right)\right]}, \nonumber
\end{eqnarray}
it is symmetric in three its arguments and changes its sign under change of any of its arguments to the conjugate one: $\Delta_{\phi_{1}} \to d - \Delta_{\phi_{1}}$ etc.

We also introduced for brevity in the last line of (\ref{24}):

\begin{equation}
\label{27}
F(\Delta) = \frac{\Gamma(\Delta) \, \Gamma(d - \Delta)}{\Gamma\left(\Delta - \frac{d}{2}\right) \, \Gamma\left(\frac{d}{2} - \Delta\right)}; \, \, \,  F_{(d=4)} = (\Delta - 1)(\Delta - 2)^{2}(\Delta - 3).
\end{equation}

\vspace{0.3cm}

Expressions for ${\bf {\cal R}}$ (\ref{26}) and $F$ (\ref{27}) are received in (\ref{24}) with account of formulas for $\delta_{ij}$, $B$, $A$ and also $C_{\Delta}$ hidden in $B$, given in (\ref{13}), (\ref{14}), (\ref{17}) and (\ref{6}) correspondingly.

Thus substitution of (\ref{24}) to (\ref{23}) gives finally:

\begin{eqnarray}
\label{28}
{\cal {\widetilde M}}^{{\rm 2pt \, bub}}_{\Delta_{\phi_{1}}|\Delta_{\phi_{2}}\Delta_{\phi_{3}}}({\vec x}_{1}, {\vec x}_{2}) = \frac{C_{\Delta_{\phi_{1}}}}{P_{12}^{\Delta_{\phi_{1}}}} \, \cdot \, \frac{g_{R}^{2}}{F(\Delta_{\phi_{1}})} \, \Biggl[{\bf {\cal R}}(\Delta_{\phi_{1}}, \Delta_{\phi_{2}}, \Delta_{\phi_{3}}) -   \qquad  \nonumber
\\
\\
- \int_{-\infty}^{+\infty} \frac{i\nu \, d\nu}{2 \pi} \, \frac{{\bf {\cal R}}(\Delta_{\phi_{1}}, \frac{d}{2} + i\nu, \Delta_{\phi_{3}})}{[\nu^{2} + (\Delta_{\phi_{2}} - \frac{d}{2})^{2}]} -\int_{-\infty}^{+\infty} \frac{i\nu \, d\nu}{2 \pi} \, \frac{{\bf {\cal R}}(\Delta_{\phi_{1}}, \Delta_{\phi_{2}}, \frac{d}{2} + i\nu)}{[\nu^{2} + (\Delta_{\phi_{3}} - \frac{d}{2})^{2}]}\, \Biggr]. \nonumber
\end{eqnarray}

\vspace{0.3cm}

Coefficient ${\bf {\cal R}}$ (\ref{26}) is the main object for us. In particular for $d = 4$ when one of its arguments is changed to $2 + i\nu$ (like for $\Delta_{\phi_{2}}$ in the second term in square brackets in the RHS of (\ref{28}) and for $\Delta_{\phi_{3}}$ in the third term) its dependence on $\nu$ is given by the elementary functions (see \cite{Ryzhik}, formula 8.332.4):

\begin{eqnarray}
\label{29}
{\bf {\cal R}}_{(d =4)}(\Delta_{\phi_{1}}, 2 + i\nu, \Delta_{\phi_{3}}) = \frac{\sin \pi\Delta_{\phi_{1}}}{\pi} \, \frac{\sin \pi\Delta_{\phi_{3}}}{\pi} \, \frac{\sinh \pi\nu}{i\, \pi} \, \cdot  \qquad \qquad \nonumber
\\
\\
\cdot \, \frac{\pi^{2} [\nu^{2} + (\Delta_{\phi_{1}} - \Delta_{\phi_{3}})^{2}]}{2\, [\cosh \pi\nu - \cos\pi(\Delta_{\phi_{1}} - \Delta_{\phi_{3}})]} \, \frac{\pi^{2} [\nu^{2} + (\Delta_{\phi_{1}} + \Delta_{\phi_{3}} - 4)^{2}]}{2\, [\cosh \pi\nu - \cos\pi(\Delta_{\phi_{1}} + \Delta_{\phi_{3}} - 4)]} \quad \nonumber
\end{eqnarray}

\vspace{0.3cm}

Thus for $d = 4$ with account of (\ref{29}) it is obtained for the UV-finite self-energy one-loop correlator (\ref{28}):

\begin{eqnarray}
\label{30}
{\cal {\widetilde M}}^{{\rm 2pt \, bub}\,(d = 4)}_{\Delta_{\phi_{1}}|\Delta_{\phi_{2}}\Delta_{\phi_{3}}}({\vec x}_{1}, {\vec x}_{2}) = \frac{C_{\Delta_{\phi_{1}}}}{P_{12}^{\Delta_{\phi_{1}}}} \, \frac{g_{R}^{2}}{F_{(d = 4)}(\Delta_{\phi_{1}})} \, \cdot \, \Biggl[ \, {\bf {\cal R}}_{(d =4)}(\Delta_{\phi_{1}}, \Delta_{\phi_{2}}, \Delta_{\phi_{3}}) - \nonumber
\\  \nonumber
\\  \nonumber
- \, \, \, \frac{\sin \pi\Delta_{\phi_{1}} \, \sin \pi\Delta_{\phi_{3}}}{8} \, \, {\rm{\bf I}}(\Delta_{\phi_{2}} - 2, \Delta_{\phi_{1}} - \Delta_{\phi_{3}}, \Delta_{\phi_{1}} + \Delta_{\phi_{3}} - 4) - \qquad \nonumber
\\
\\
- \, \frac{\sin \pi\Delta_{\phi_{1}} \, \sin \pi\Delta_{\phi_{2}}}{8} \, \, {\rm{\bf I}}(\Delta_{\phi_{3}} - 2, \Delta_{\phi_{1}} - \Delta_{\phi_{2}}, \Delta_{\phi_{1}} + \Delta_{\phi_{2}} - 4) \Biggl], \quad \qquad \nonumber
\end{eqnarray}
where we introduced definite integral:

\begin{equation}
\label{31}
{\rm{\bf I}}(a, b, c) = \int_{-\infty}^{+ \infty} \, \frac{\nu \sinh \pi\nu \, d\nu}{\nu^{2} + a^{2}} \, \frac{\nu^{2} + b^{2}}{[\cosh \pi\nu - \cos \pi b]}\ \, \frac{\nu^{2} + c^{2}}{[\cosh \pi\nu - \cos \pi c]}.
\end{equation}

\section{$O(N)$-symmetric model: successful "hunting for numbers"}

\qquad As it was noted in the Introduction the AdS/CFT version of the "old" conformal bootstrap is obtained if coordinates $X_{1,2}$, $X$, $Y$ in general formula (\ref{1}) are considered as the bulk coordinates in $AdS_{d + 1}$, and $X_{1,2}$ are placed at the horizon with appropriate normalization like in (\ref{6}), (\ref{7}). This procedure transforms LHS of (\ref{1}) into elementary boundary-to-boundary conformal correlator (\ref{7}) whereas RHS of (\ref{1}) transforms to the UV-divergent 2-point one-loop self-energy correlator; we use its UV-finite redefined form (\ref{20}) built with general approach (\ref{3}) of the double-trace from UV to IR deformation. Thus "old" conformal bootstrap in the AdS/CFT context looks as:

\begin{equation}
\label{32}
\frac{C_{\Delta_{\phi_{1}}}}{P_{12}^{\Delta_{\phi_{1}}}} \,  = \,  {\cal {\widetilde M}}^{{\rm 2pt \, bub}}_{\Delta_{\phi_{1}}|\Delta_{\phi_{2}}\Delta_{\phi_{3}}}({\vec x}_{1}, {\vec x}_{2}).
\end{equation} 

Space dependence of ${\widetilde M}$ is singled out in front of (\ref{28}) and it reduces in (\ref{32}) with the same dependence of the LHS of (\ref{32}). Two more bootstrap equations are obtained from (\ref{32}) by the permutation of fields. Thus we have three spectral equations for four unknown variables: $\Delta_{\phi_{1}}$, $\Delta_{\phi_{2}}$, $\Delta_{\phi_{3}}$ and $g_{R}^{2}$.

The most consistent way to get the missing fourth equation for coupling constant would be to require the validity of the vertex "old" bootstrap equation that in the AdS/CFT context means equating of 3-point vertex (\ref{12}) to the one-loop triangle Witten diagram, redefined according to prescription (\ref{3}). This bootstrap equation is easy to put down, but it is difficult to work it out.

Another option is to reduce the number of unknown variables with fixing conformal dimension of one of the fields. This will be done below.

Let us look at the $O(N)$ symmetric model of $N$ scalar fields $\psi_{k}$ with quartic interaction term $\sim (\Sigma_{k}\psi_{k}^{2})^{2}$ on the $AdS_{d + 1}$ background. This model may be always reduced to triple interaction

\begin{equation}
\label{33}
L_{int} = g \, \sigma(Z)\,\Sigma_{k}\psi_{k}^{2}(Z)
\end{equation}
with the introduction of the auxiliary Hubbard-Stratonovich field $\sigma(Z)$. 

Thus we consider theory of $N + 1$ scalar fields with interaction (\ref{33}) where $N$ fields $\psi_{k}$ have one and the same conformal dimension $\Delta_{\psi} > d/2$ and conformal dimension $\Delta_{\sigma}$ of the Hubbard-Stratonovich field is given. In what follows the Hubbard-Stratonovich field is considered to be conformally invariant in $AdS_{d + 1}$, that is:

\begin{equation}
\label{34}
\Delta_{\sigma} = \Delta^{\rm conf}_{\sigma} = \frac{d}{2} + \frac{1}{2} \, \, \, (= \frac{5}{2} \,\, for \, \, d = 4)
\end{equation}
(we take IR-option (\ref{5}) of $\Delta_{\psi}$, $\Delta_{\sigma}$ when spectral representation (\ref{8}) of intermediate Green functions is valid).

Then there are $N$ identical bootstrap Eq-s (\ref{32}) written for fields $\psi_{k}$ and one more Eq. (\ref{32}) for field $\sigma$ where the RHS must be multiplied by $N$ since in (\ref{33}) field $\sigma(Z)$ interacts with every of fields $\psi_{k}$. To write down these spectral equations it is sufficient to put in (\ref{30}), (\ref{26}) $\Delta_{\phi_{1}} = \Delta_{\phi_{2}} = \Delta_{\psi}$ and $\Delta_{\phi_{3}} = \Delta_{\sigma}$ (\ref{34}).

In what follows we consider $d = 4$. Thus there are two bootstrap spectral equations obtained from (\ref{32}), (\ref{30}), (\ref{33}). One for field $\psi$:

\begin{eqnarray}
\label{35}
1 \, = \, \frac{g_{R}^{2}}{F_{(d = 4)}(\Delta_{\psi})} \, \cdot \, \Biggl[ \, {\bf {\cal R}}_{(d =4)}(\Delta_{\psi}, \Delta_{\psi}, \Delta_{\sigma}) - \qquad  \qquad   \nonumber
\\  \nonumber
\\  \nonumber
- \, \, \, \frac{\sin \pi\Delta_{\psi} \, \sin \pi\Delta_{\sigma}}{8} \, \, {\rm{\bf I}}(\Delta_{\psi} - 2, \Delta_{\psi} - \Delta_{\sigma}, \Delta_{\psi} + \Delta_{\sigma} - 4) - \qquad \nonumber
\\
\\
- \, \frac{(\sin \pi\Delta_{\psi})^{2}}{8} \, \, {\rm{\bf I}}(\Delta_{\sigma} - 2, 0, 2 \Delta_{\psi} - 4) \Biggl], \quad \qquad \qquad \qquad \nonumber
\end{eqnarray}
and the other one for field $\sigma$:

\begin{eqnarray}
\label{36}
1 \, = \, N \, \frac{g_{R}^{2}}{F_{(d = 4)}(\Delta_{\sigma})} \, \cdot \, \Biggl[ \, {\bf {\cal R}}_{(d =4)}(\Delta_{\psi}, \Delta_{\psi}, \Delta_{\sigma}) - \qquad  \qquad   \nonumber
\\ 
\\ 
- \, \, \, 2 \, \frac{\sin \pi\Delta_{\sigma} \, \sin \pi\Delta_{\psi}}{8} \, \, {\rm{\bf I}}(\Delta_{\psi} - 2, \Delta_{\sigma} - \Delta_{\psi}, \Delta_{\psi} + \Delta_{\sigma} - 4)\Biggr], \qquad \nonumber
\end{eqnarray}
${\bf {\cal R}}_{(d =4)}(\Delta_{\psi}, \Delta_{\psi}, \Delta_{\sigma})$, $F_{(d = 4)}(\Delta_{\psi})$, ${\rm{\bf I}}(a,b,c)$ see in (\ref{26}), (\ref{27}), (\ref{31}).

After elimination here $g_{R}^{2}$ and substitution of $\Delta_{\sigma}$ from (\ref{34}) (for $d = 4$) the looked for spectral equation for $\Delta_{\psi}$ is obtained; it is convenient to put it down for variable $\lambda$:

\begin{equation}
\label{37}
\lambda = \Delta_{\psi} - 2 > 0; \, \, \, \lambda < 1.
\end{equation}
Here $\lambda > 0$ since $\Delta_{\psi} > d/2 = 2$; inequality $\lambda < 1$ is the unitarity bound demand. 

Finally taking into account that according to (\ref{26}), (\ref{34}), (\ref{37})

\begin{equation}
\label{38}
{\bf {\cal R}}_{(d = 4)}\left(\lambda + 2, \lambda + 2, \frac{5}{2}\right) = \frac{\pi}{8}\, \left(\lambda^{2} - \frac{1}{16}\right) \, \frac{1 - \cos 2\pi\lambda}{\cos 2\pi\lambda},
\end{equation}
and that according to (\ref{27}) $F_{(d = 4)}(\lambda + 2) = \lambda^{2}(\lambda^{2} - 1)$ and $F_{(d = 4)}(5/2) = - 3/16$ the following spectral equation for $\lambda$ (\ref{37}) is obtained from (\ref{34})-(\ref{36}), (\ref{38}):

\begin{eqnarray}
\label{39}
2\pi \sin\pi\lambda \Bigg[1 + N \frac{16}{3} \lambda^{2}(\lambda^{2} - 1)\Bigg]  \left(\lambda^{2} - \frac{1}{16}\right)  -  \sin\pi\lambda \, \cos2\pi\lambda \, {\rm{\bf I}}\left(\frac{1}{2}, 0, 2 \lambda \right) - \nonumber
\\
\\
- \cos2\pi\lambda \, \Bigg[1 + 2 N \frac{16}{3} \lambda^{2}(\lambda^{2} - 1)\Bigg] \, \, {\rm{\bf I}}\left(\lambda, \lambda - \frac{1}{2}, \lambda + \frac{1}{2}\right) = 0,  \qquad  \qquad \nonumber
\end{eqnarray}
where integral ${\rm{\bf I}}$ is defined in (\ref{31}).

For every $N = 1, 2, 3, 4$ there are three roots of Eq. (\ref{39}) obeying unitarity bound demand $0 < \lambda < 1$, the values of $g_{R}^{2}$ (in units of the proper powers of the AdS curvature) calculated from (\ref{35}) or (\ref{36}) corresponding to every of these roots are also shown below:

\begin{eqnarray}
\label{40}
N = 1: \, \lambda = \, \,\, 0.500 \, (g_{R}^{2} = 0.75); \, \, \, 0.875 \, (g_{R}^{2} = - 2.74); \,\, \, 0.965 \, (g_{R}^{2} = 36.2); \nonumber
\\  \nonumber
\\ \nonumber
N = 2: \, \lambda = \, \, \, 0.296 \, (g_{R}^{2} = 0.72); \, \, \, 0.928 \, (g_{R}^{2} = - 16.9); \,\, \, 0.978 \, (g_{R}^{2} = 18.7); \nonumber
\\
\\
N = 3: \, \lambda = \, \, \, 0.227 \, (g_{R}^{2} = 0.64); \, \, \, 0.936 \, (g_{R}^{2} = - 17.9); \,\, \, 0.985 \, (g_{R}^{2} = 14.0);  \nonumber
\\ \nonumber
\\ \nonumber
N = 4: \, \lambda = \, \, \, 0.189 \, (g_{R}^{2} = 0.57); \, \, \, 0.938 \, (g_{R}^{2} = - 19.6); \,\, \, 0.988 \, (g_{R}^{2} = 12.5); \, \nonumber
\end{eqnarray}
these values of $\lambda$ correspond to conformal dimensions $2 < \Delta^{IR}_{\psi} < 3$ (see (\ref{37})).

Surely for every positive solution $\lambda > 0$ (\ref{40}) of the "old" conformal bootstrap equations (\ref{32}) there exists negative solution of the same modulus that corresponds to the conjugate conformal dimension $\Delta^{UV}_{\psi} = 4 - \Delta^{IR}_{\psi} = 2 - |\lambda|$ . However it would be mistake just to reverse the sign of $\lambda$ in spectral equation (\ref{39}). Spectral equation identical to (\ref{39}) is obtained for $|\lambda|$, that is for "conjugate sector", if in general bootstrap equations (\ref{32}) conformal dimensions of every "external" tail is taken "UV" that is less than $d/2$. For model under consideration and for $d = 4$ this means that in Eq. (\ref{32}) written down for "external" field $\psi$ it is necessary to replace $\Delta_{\phi_{1}} \to 4 - \Delta_{\psi}$ ($\Delta_{\psi} > 2$) and in Eq. (\ref{32}) for "external" field $\sigma$ we must take $\Delta_{\sigma} = 3/2$, while conformal dimensions of "intermediate" fields should remain "IR" when spectral representations (\ref{8}) of Green functions forming the bubble are valid. With such a substitutions in (\ref{32}) the RHS of bootstrap equations (\ref{35}), (\ref{36}) change sign, which does not change the spectral equation (\ref{39}) written for $|\lambda|$, but replaces $g_{R}^{2} \to - g_{R}^{2}$.

\section{Conclusion}

\qquad Solutions (\ref{40}) are $O(N)$ symmetric, they were obtained under the assumption of the coincidence of conformal dimensions of all fields $\psi_{k}$. The immediate task for future would be to look at the possibility of spontaneous $O(N)$ symmetry breaking in the model (\ref{33}) when every of fields $\psi_{k}$ is equipped with its own conformal dimension $\Delta_{\psi_{k}}$ and asymmetric solutions of self-consistent bootstrap equations (\ref{32}) must be found. In particular this means in case of $d = 4$ that bootstrap Eq. (\ref{35}) is valid for every $\Delta_{\psi_{k}}$ whereas in (\ref{36}) instead of factor $N$ in the RHS there will be a sum over $k$ of functions on $\Delta_{\psi_{k}}$ standing in the RHS of (\ref{36}).

The simplest case of the interacting bulk scalar fields is considered in the paper, and the physical meaning of solutions (\ref{40}) is rather vague. However many questions of modern physics are connected with the fermions of spin $1/2$, and the point is that "flavors" mass hierarchy, which is still a mystery, may be explained in principle in frames of the two-brane Randall-Sundrum model \cite{Randall} when some natural ("twisted") boundary conditions are imposed on the bulk spinor fields and for certain bulk masses of these fields (see e.g. \cite{Neubert}, \cite{Pomarol}). Thus calculation of fermions' bulk masses (that is of conformal dimensions) in frames of the proposed approach of the "old" conformal bootstrap may open the way to solving the problem of the fermion mass hierarchy. To generalize the approach of the present paper to fields of spin 1/2 in frames of Yukawa model or QED on $AdS_{5}$ may be one more immediate task for future.

\section*{Acknowledgments} Author is grateful to Ruslan Metsaev for fruitful discussions and to participants of the seminar in the Theoretical Physics Department of P.N. Lebedev Physical Institute for stimulating questions.

\end{document}